\documentclass[10pt,letterpaper,twocolumn,american,nofootinbib]{revtex4}
\usepackage{ae}
\usepackage{aecompl}
\usepackage[T1]{fontenc}
\usepackage[latin1]{inputenc}
\usepackage{amsmath}
\usepackage{graphicx}

\makeatletter
\usepackage{psfrag}
\usepackage{babel}
\makeatother
\begin{document}

\title{Drawing conformal diagrams for a fractal landscape }

\author{Sergei Winitzki}

\affiliation{Department of Physics, Ludwig-Maximilians University, Theresienstr.~37, 80333
Munich, Germany}

\begin{abstract}
Generic models of cosmological inflation and the recently proposed scenarios
of a recycling universe and the string theory landscape predict spacetimes whose
global geometry is a stochastic, self-similar fractal. To visualize the complicated
causal structure of such a universe, one usually draws a conformal (Carter-Penrose)
diagram. I develop a new method for drawing conformal diagrams, applicable to
arbitrary 1+1-dimensional spacetimes. This method is based on a qualitative
analysis of intersecting lightrays and thus avoids the need for explicit transformations
of the spacetime metric. To demonstrate the power and simplicity of this method,
I present derivations of diagrams for spacetimes of varying complication. I
then apply the lightray method to three different models of an eternally inflating
universe (scalar-field inflation, recycling universe, and string theory landscape)
involving the nucleation of nested asymptotically flat, de Sitter and/or anti-de
Sitter bubbles. I show that the resulting diagrams contain a characteristic
fractal arrangement of lines. 
\end{abstract}
\maketitle

\section{Introduction}

Generic models of cosmological inflation predict that thermalization will never
be reached everywhere in the universe. This phenomenon, termed \textsl{eternal
inflation}, was first analyzed for scalar-field inflationary scenarios~\cite{Vil83,Lin86}.
The resulting spacetime is inhomogeneous since some regions will still undergo
inflation while other regions have thermalized. Inhomogeneities of this type
occur at distances much larger than the Hubble scale $H^{-1}$. With eternal
inflation, the proper volume of the non-thermalized (inflating) domain grows
with time and forms a self-similar lacunary fractal, first studied in Ref.~\cite{VilAry89}.
An increasing number of regions thermalize at progressively later times (this
statement is independent of the choice of an equal-time slicing), and moreover
it can be shown that there exist infinitely many comoving geodesics that never
enter any thermalized regions~\cite{Win02}. 

Besides scalar field-driven inflation, there exist other cosmological scenarios
with a similar stochastic and fractal geometry of the spacetime. These are for
instance the models of a \textsl{recycling universe}~\cite{GarVil98} and the
\textsl{string theory landscape}~\cite{landscape}. In these scenarios, various
de Sitter (or anti-de Sitter) regions can nucleate as bubbles within the background
de Sitter spacetime and within other bubbles. It is an interesting challenge
to visualize the causal structure of such a spacetime. In particular, the geometry
of an asymptotic future infinity is relevant to considerations of the holographic
principle~\cite{tHo93,Susskind,Bousso}. 

The purpose of this paper is to construct conformal diagrams (also called \textsl{Carter-Penrose
diagrams}) for spacetimes arising from models of eternal inflation. All such
spacetimes consist of approximately homogeneous Hubble-size regions expanding
at different rates and forming a fractal structure on super-Hubble scales. I
describe a new method of drawing conformal diagrams for spacetimes where the
metric cannot be obtained in closed form. This method is based on a qualitative
analysis of the asymptotic behavior of lightrays in the given spacetime and
does not require deriving explicit conformal transformations of the metric.
I illustrate the new method on some standard as well as novel examples. I then
apply this method to construct conformal diagrams for various models of an eternally
inflating universe. The models of interest are: a generic inflation producing
either matter-dominated or dark energy-dominated domains; a recycling universe
that generates a never-ending sequence of nested thermalized and inflating (de
Sitter) domains; and the string theory landscape which includes recycling and
also admits anti-de Sitter bubbles ending in a singularity. I show that conformal
diagrams representing the resulting spacetimes can be drawn using a random fractal
arrangement of lines. I present computer-simulated conformal diagrams that help
visualize the causal structure of such spacetimes. In particular, it becomes
apparent that these spacetimes possess an infinite number of points representing
different causally disconnected null and timelike conformal infinities.

\section{Drawing conformal diagrams}

After recapitulating the standard construction of conformal diagrams, I shall
develop a new, easier method that will be used throughout the paper.

\subsection{Standard procedure}

A conformal diagram is defined for a 1+1-dimensional spacetime with a given
line element $g_{ab}dx^{a}dx^{b}$. The coordinates $x^{a}$ (where $a=0,1$)
must cover the entire spacetime, and several sets of overlapping coordinate
patches may be used if necessary. The standard construction of the conformal
diagram may be formulated as follows (see e.g.~\cite{BirDav82}, chapter~3).
One first finds a change of coordinates $x\rightarrow\tilde{x}$ such that the
new variables $\tilde{x}^{a}$ have a \emph{finite} range of variation; the
components of the metric change according to $g_{ab}(x)dx^{a}dx^{b}=\tilde{g}_{ab}(\tilde{x})d\tilde{x}^{a}d\tilde{x}^{b}$.
One then chooses a conformal transformation of the metric (in the new coordinates),
\begin{equation}
\tilde{g}_{ab}\rightarrow\gamma_{ab}\left(\tilde{x}\right)=\Omega^{2}\left(\tilde{x}\right)\tilde{g}_{ab},\quad\Omega\left(\tilde{x}\right)\neq0,\label{eq:g transf}\end{equation}
such that the new metric $\gamma_{ab}$ is flat, i.e.~has zero curvature. A
suitable function $\Omega(\tilde{x})$ always exists because all two-dimensional
metrics are conformally flat. The new metric $\gamma_{ab}$ describes an unphysical,
auxiliary flat spacetime. Since the metric $\gamma_{ab}$ is flat, a further
change of coordinates $\tilde{x}\rightarrow\tilde{\tilde{x}}$ (the new variables
$\tilde{\tilde{x}}$ again having a finite extent) can be found to transform
$\gamma_{ab}$ explicitly into the Minkowski metric $\eta_{ab}$, \begin{equation}
\gamma_{ab}\left(\tilde{x}\right)d\tilde{x}^{a}d\tilde{x}^{b}=\eta_{ab}d\tilde{\tilde{x}}^{a}d\tilde{\tilde{x}}^{b},\quad\eta_{ab}\equiv\textrm{diag}\left(1,-1\right).\label{eq:omega eta}\end{equation}
For brevity we incorporate all the required coordinate changes into one, $x\rightarrow\tilde{x}(x)$,
and summarize the transformation of the metric as \begin{equation}
\Omega^{2}\left(x\right)g_{ab}dx^{a}dx^{b}=\eta_{ab}d\tilde{x}^{a}d\tilde{x}^{b}.\label{eq:g transf short}\end{equation}
 Thus the new coordinates $\tilde{x}^{a}$ map the initial spacetime onto a
\emph{finite} domain within a 1+1-dimensional Minkowski plane. This finite domain
is a \textsl{conformal diagram} of the initial (physical) 1+1-dimensional spacetime.
The diagram is drawn on a sheet of paper which implicitly carries the fiducial
Minkowski metric $\eta_{ab}$, the vertical axis usually representing the timelike
coordinate $\tilde{x}^{0}$.

The coordinate and conformal transformations severely distort the geometry of
the spacetime since they bring infinite spacetime points to finite distances
in the diagram plane. So one cannot expect in general that straight lines in
the diagram correspond to geodesics in the physical spacetime. However, it is
well-known that straight lines drawn at $45^{\circ}$ angles in a conformal
diagram represent null geodesics in the physical spacetime. This follows from
the fact that any null trajectory $x^{a}(\tau)$ in 1+1 dimensions, i.e.~any
solution of \begin{equation}
g_{ab}\dot{x}^{a}\dot{x}^{b}=0,\quad\dot{x}^{a}\equiv\frac{dx^{a}}{d\tau},\label{eq:g x x}\end{equation}
is necessarily a geodesic (this is not true in higher dimensions), and Eq.~(\ref{eq:g x x})
is invariant under conformal transformations of the metric. By drawing lightrays
emitted from various points in the diagram, one can illustrate the causal structure
of the spacetime.

\begin{figure}
\begin{center}\psfrag{x0}{$\tilde{x}^0$}\psfrag{x1}{$\tilde{x}^1$}

\psfrag{1}{$1$}\psfrag{A}{$A$}\psfrag{B}{$B$}\psfrag{0}{$0$}\psfrag{-1}{$-1$}\includegraphics[%
  width=2in]{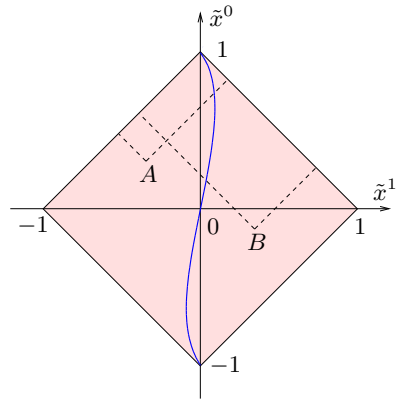}\end{center}

\caption{A conformal diagram of the 1+1-dimensional Minkowski spacetime. Dashed lines
show lightrays emitted from points $A,B$. The curved line is the trajectory
of an inertial observer moving with a constant velocity, $x^{1}=0.3x^{0}$.
\label{cap:Minkowski}}
\end{figure}

A textbook example is the conformal diagram for the flat Minkowski spacetime
with the metric $g_{ab}\equiv\eta_{ab}$. Calculations are conveniently done
in the lightcone coordinates \begin{equation}
u\equiv x^{0}-x^{1},\; v\equiv x^{0}+x^{1},\;\eta_{ab}dx^{a}dx^{b}=du\, dv,\end{equation}
 and a suitable coordinate transformation is \begin{equation}
\tilde{u}=\tanh u,\;\tilde{v}=\tanh v,\; du\, dv=\frac{d\tilde{u}\, d\tilde{v}}{\left(1-\tilde{u}^{2}\right)\left(1-\tilde{v}^{2}\right)}.\label{eq:Mink lightcone}\end{equation}
The new coordinates $\tilde{u},\tilde{v}$ extend from $-1$ to $1$. Multiplying
the metric by the conformal factor $\Omega^{2}(\tilde{u},\tilde{v})\equiv\left(1-\tilde{u}^{2}\right)\left(1-\tilde{v}^{2}\right)$,
we obtain the fiducial spacetime,\begin{align}
 & \Omega^{2}du\, dv=d\tilde{u}\, d\tilde{v}=\eta_{ab}d\tilde{x}^{a}d\tilde{x}^{b},\\
 & \tilde{u}\equiv\tilde{x}^{0}-\tilde{x}^{1},\quad\tilde{v}\equiv\tilde{x}^{0}+\tilde{x}^{1}.\end{align}
The new coordinates $\tilde{x}^{a}$ have a finite extent, namely $\left|\tilde{x}^{0}\pm\tilde{x}^{1}\right|<1$,
and the resulting diagram has a diamond shape shown in Fig.~\ref{cap:Minkowski}.
To appreciate the distortion of the spacetime geometry, we can draw the worldline
of an inertial observer moving with a constant velocity. Note that the angle
at which this trajectory enters the endpoints depends on the chosen conformal
transformation and thus cannot serve as an indication of the observer's velocity.

There is no analog of conformal diagrams for general 3+1-dimensional spacetimes.
Nevertheless, in many cases a 3+1-dimensional spacetime can be adequately represented
by a suitable 1+1-dimensional slice, at least for the purpose of qualitative
illustration. For instance, a spherically symmetric spacetime is visualized
as the $\left(t,r\right)$ half-plane ($r\geq0$) where each point stands for
a 2-sphere of radius $r$. A conformal diagram is then drawn for the reduced
1+1-dimensional spacetime.

The reduced conformal diagram is meaningful only if the null geodesics in the
1+1-dimensional section are also geodesics in the physical 3+1-dimensional spacetime.
In this case, the 1+1-dimensional section can be visualized as the set of events
accessible to an observer who sends and receives signals only along a fixed
spatial direction. Then a conformal diagram provides information about the causal
structure of spacetime along this line of sight.

\subsection{Method of lightrays}

The standard construction of conformal diagrams involves an explicit transformation
of the metric to new coordinates that have a finite extent, and usually a further
transformation to bring the metric to a manifestly conformally flat form. Finding
these transformations requires a certain ingenuity. If the spacetime manifold
is covered by several coordinate patches, a different transformation must be
used in each patch. However, a conformal diagram typically consists of just
a few lines and one would expect that the required computations should not be
so cumbersome. 

I now describe a method of drawing conformal diagrams that avoids the need for
performing explicit transformations of the metric. The method is based on a
qualitative analysis of intersections of lightrays. This approach is particularly
suitable for the analysis of stochastic spacetimes encountered in models of
eternal inflation. Such spacetimes have no symmetries and their metric is not
known in closed form, so one cannot apply the standard construction of conformal
diagrams. 

Another motivation for the new method is the apparent redundancy involved in
the standard method. It is clear that the transformations used in the standard
construction are not unique. For instance, one may replace the lightcone coordinates
in Eq.~(\ref{eq:Mink lightcone}) by\begin{equation}
\tilde{u}\rightarrow f\left(\tilde{u}\right),\quad\tilde{v}\rightarrow g\left(\tilde{v}\right),\label{eq:u v change}\end{equation}
 where $f,g$ are arbitrary monotonic, bounded, and continuous functions. The
shape of the diagram will vary with each possible choice of the transformations,
but all resulting diagrams are equivalent in the sense that they contain the
same information about the causal structure of the spacetime. One thus expects
to be able to extract this information without involving specific explicit transformations
of the coordinates and the metric.

The crucial observation is that this information is unambiguously represented
by the geometry and topology of lightrays and their intersections. I shall now
develop this idea into a self-contained approach to drawing conformal diagrams
that does not involve explicit transformations.

\subsubsection{New definition of conformal diagrams}

A conformal diagram is a figure in the fiducial Minkowski plane satisfying certain
conditions, and I first formulate a definition of a conformal diagram in terms
of such conditions. A constructive procedure for drawing conformal diagrams
will be presented subsequently.

A finite open domain of the plane is a \textsl{conformal diagram} of a given
1+1-dimensional spacetime $S$ if there exists a one-to-one correspondence between
all maximally extended lightrays in $S$ and all straight line segments drawn
at $45^{\circ}$ angles within the domain of the diagram. This correspondence
must be \textsl{intersection-preserving}, i.e.~any two lightrays intersect
in the physical spacetime exactly as many times as the corresponding lines intersect
in the diagram. It is assumed that all null geodesics in the spacetime $S$
are either infinitely extendible or end at singularities or at explicitly introduced
spacetime boundaries. Similarly, the straight line segments drawn at $45^{\circ}$
angles in a conformal diagram must be limited only by the boundary of the diagram.
Note that there are only two spatial directions in a 1+1-dimensional spacetime
$S$, and that two lightrays emitted in the same direction cannot intersect.

For example, the diamond $\left|\tilde{x}^{0}\pm\tilde{x}^{1}\right|<1$ is
a conformal diagram for the flat spacetime due to the intersection-preserving
one-to-one correspondence of null lines $\tilde{x}^{0}\pm\tilde{x}^{1}=\textrm{const}$
in the diagram and lightrays $x^{0}\pm x^{1}=\textrm{const}$ in the physical
spacetime.

For spacetimes having a nontrivial topology, appropriate topological features
need to be introduced also into the fiducial Minkowski plane. At this point
I do not consider such cases.

\begin{figure}
\begin{center}\psfrag{x0}{${x}^0$}\psfrag{x1}{${x}^1$}

\psfrag{y0}{$\tilde{x}^0$}\psfrag{y1}{$\tilde{x}^1$}

\psfrag{correspondence}{correspondence}\includegraphics[%
  width=3in]{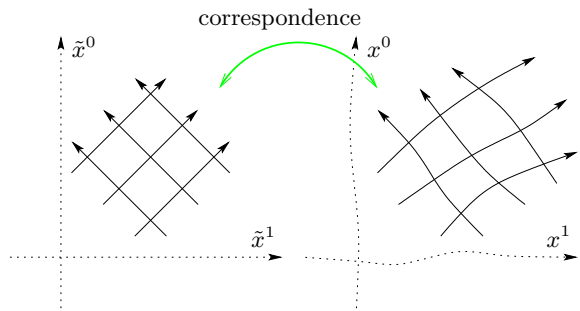}\end{center}

\caption{The local correspondence between straight lines drawn at $45^{\circ}$ angles
in the conformal diagram (left) and lightrays in the physical spacetime (right).
\label{cap:correspondence}}
\end{figure}

I shall now demonstrate the equivalence of the proposed definition to the standard
procedure for drawing conformal diagrams. It suffices to find a conformal transformation
of the form~(\ref{eq:g transf short}) in some neighborhood of an arbitrary
(nonsingular) spacetime point. Given a diagram with an intersection-preserving
correspondence of lightrays, we can introduce local lightcone coordinates $u,v$
in the diagram such that the null geodesics are locally the lines $u=\textrm{const}$
or $v=\textrm{const}$. By assumption, each null geodesic uniquely corresponds
to a lightray in the physical spacetime. Since the correspondence is intersection-preserving,
the local configuration of the null geodesics in the physical spacetime can
be visualized as in Fig.~\ref{cap:correspondence}. Hence the local lightcone
coordinates $u,v$ become well-defined local coordinates in the physical spacetime,
and again the null geodesics are the lines $u=\textrm{const}$ or $v=\textrm{const}$.
On the other hand, these null geodesics must be solutions of Eq.~(\ref{eq:g x x}),
therefore\begin{equation}
g_{uu}\dot{u}^{2}+2g_{uv}\dot{u}\dot{v}+g_{vv}\dot{v}^{2}=0\,\textrm{ if }\,\dot{u}=0\textrm{ or }\dot{v}=0.\end{equation}
 It follows that $g_{uu}=g_{vv}=0$. Thus the metric in the local lightcone
coordinates is of the form $g_{ab}dx^{a}dx^{b}=2g_{uv}(u,v)du\, dv$ which is
explicitly conformally flat. This demonstrates the existence of a local conformal
transformation bringing the physical metric $g_{ab}$ into the fiducial Minkowski
metric $du\, dv=\eta_{ab}d\tilde{x}^{a}d\tilde{x}^{b}$ in the diagram.

Before presenting examples, I comment on the proposed definition of conformal
diagrams. The definition may appear to be too broad, allowing many geometric
shapes to represent the same spacetime. However, the old procedure also does
not specify a particular conformal transformation of the metric and in effect
admits precisely as much freedom. According to the new definition, any two different
conformal diagrams of the same spacetime are equivalent in the sense that all
the lightrays in those two diagrams will be in an intersection-preserving one-to-one
correspondence. In the old language, there must exist a conformal transformation
bringing one diagram into the other. Equivalent diagrams carry identical information
about the causal structure of the physical spacetime. In the next sections I
shall give examples illustrating the arbitrary and the necessary choices involved
in drawing conformal diagrams.

A conformal diagram delivers information mainly through the shape of its boundary
line. The boundary of a conformal diagram generally contains points representing
a spacelike, timelike, or null infinity, and points belonging to explicit boundaries
of the spacetime manifold (e.g.~singularities) where lightrays end in the physical
spacetime. The latter boundaries will be called \textsl{physical boundaries}
to distinguish them from putative \textsl{infinite boundaries} whose points
do not correspond to any points in the physical spacetime. (The definition of
conformal diagrams contains the requirement that the diagram domain be topologically
open, and so the boundary points are not supposed to belong to the diagram.)
The infinite boundary is of course the most interesting feature of a diagram.

Lastly, I would like to emphasize that conformal diagrams can be drawn not only
for geodesically complete spacetimes but also for {}``artificially incomplete''
spacetimes, i.e.~for selected subdomains of larger manifolds. In fact such
{}``artificially incomplete'' spacetimes are often needed in cosmological
applications. Examples are a description of a collapsing star using a subdomain
of the Kruskal spacetime and a description of an inflationary universe using
a subdomain of the de Sitter spacetime.

\subsubsection{Minkowski spacetime\label{sub:Minkowski-spacetime}}

The new definition merely lists the conditions to be satisfied by a conformal
diagram. Based on these conditions, I shall now develop a practical procedure
for drawing the diagrams, using the Minkowski spacetime as the first example. 

\begin{figure}
\begin{center}\psfrag{A}{$A$}\psfrag{B}{$B$}\psfrag{C}{$C$}\psfrag{D}{$D$}

\psfrag{E}{$E$}

\psfrag{L}{Cauchy line $\tilde {L}$}

\psfrag{?}{???}\includegraphics[%
  width=2.5in]{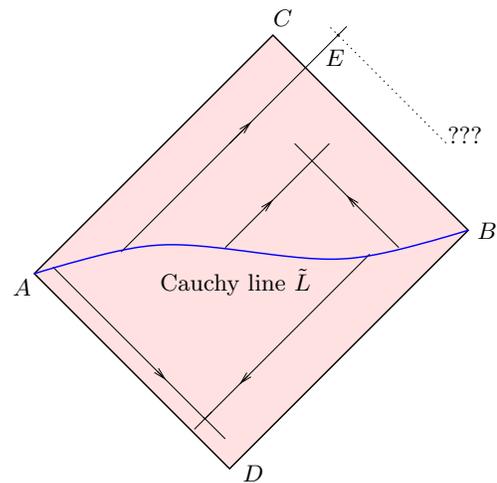}\end{center}

\caption{Construction of the conformal diagram for the Minkowski spacetime. The point
$E$ cannot belong to the diagram because there are no lightrays intersecting
at $E$. \label{cap:build Mink}}
\end{figure}

We begin by choosing a Cauchy surface in the physical spacetime. In 1+1 dimensions,
a Cauchy surface is a line $L$ such that intersections of lightrays emitted
from $L$ entirely cover the part of the spacetime to the future of $L$. In
the Minkowski spacetime, we may choose the line $x^{0}=0$ as the Cauchy surface
$L$. The image of the Cauchy surface $L$ in the conformal diagram must be
a finite curve $\tilde{L}$ from which lightrays can be emitted in both spatial
directions. Therefore the slope of the curve $\tilde{L}$ must not exceed $\pm45^{\circ}$
but otherwise $\tilde{L}$ may be drawn arbitrarily, e.g.~as the curve $AB$
in Fig.~\ref{cap:build Mink}. The endpoints $A,B$ represent a spatial infinity
in the two directions.

In the Minkowski spacetime, two lightrays emitted towards each other from any
two points on $L$ will eventually intersect. Therefore the domain of the conformal
diagram must contain at least the triangular region $ABC$. On the other hand,
any point outside $ABC$, such as the point $E$ in Fig.~\ref{cap:build Mink},
cannot belong to the diagram domain because the point $E$ cannot be reached
by any left-directed lightray emitted from $L$, and we know that all points
in the Minkowski spacetime are intersection points of some lightrays. (More
formally, the existence of the point $E$ within the diagram would violate the
condition that the correspondence between lightrays is intersection-preserving.)
Therefore the future-directed part of the conformal diagram is bounded by the
lines $AC$ and $BC$.

A completely analogous consideration involving past-directed lightrays leads
to the conclusion that the past-directed part of the diagram is the region $ABD$.
Thus a possible diagram for the Minkowski spacetime is the interior of the rectangle
$ACBD$. This diagram differs from the square-shaped diagram in Fig.~\ref{cap:Minkowski}
by a (finite) conformal transformation of the form~(\ref{eq:u v change}). 

We can also ascertain that the points $C$ and $D$ are the future and the past
timelike infinity points. For instance, the point $C$ is the intersection of
the lines $AC$ and $BC$; these lines are interpreted as putative lightrays
emitted from infinitely remote points of $L$. At sufficiently late times, any
inertial observer in the Minkowski spacetime will catch lightrays emitted from
arbitrarily far points. The same holds for observers moving non-inertially as
long as their velocity does not approach that of light. Hence all trajectories
of such observers must finish at $C$.

\subsection{Cauchy surfaces and artificial boundaries\label{sub:Cauchy-surfaces-and}}

Drawing Cauchy surfaces is a convenient starting point in the construction of
conformal diagrams.

\begin{figure}
\begin{center}\psfrag{boundary}{boundary}

\psfrag{Cauchy line L}{Cauchy line $\tilde {L}$}\includegraphics[%
  width=2.5in]{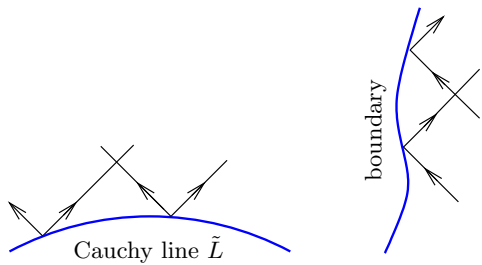}\end{center}

\caption{A Cauchy line must have a slope between $-45^{\circ}$ and $45^{\circ}$ (left).
A timelike boundary must be a curve with a slope between $45^{\circ}$ and $135^{\circ}$
(right). \label{cap:build lines}}
\end{figure}

It is clear that in any 1+1-dimensional spacetime a sufficiently small neighborhood
of a Cauchy surface has the same causal properties as the line $x^{0}=0$ in
the Minkowski plane: namely, two nearby lightrays emitted toward each other
will cross, while rays emitted from a point in opposite directions will diverge.
Therefore the line $\tilde{L}$ representing a Cauchy surface in a conformal
diagram must have a slope between $-45^{\circ}$ and $45^{\circ}$ (see Fig.~\ref{cap:build lines},
left). We shall call such lines \textsl{horizontally-directed}. Other than this,
there are no restrictions on the shape of the line $\tilde{L}$ and it may be
drawn as an arbitrary horizontally-directed curve. (In some spacetimes, one
needs to use several disconnected Cauchy surfaces, but I shall not consider
such cases here.)

Another frequently occurring feature in conformal diagrams is a \textsl{timelike
boundary}. For example, a spherically symmetric 3+1-dimensional spacetime is
usually reduced to the 1+1-dimensional $\left(r,t\right)$ plane, where $0<r<+\infty$.
From the 1+1-dimensional point of view, the line $r=0$ is an artificially introduced
timelike boundary that can absorb and emit lightrays. The local geometry of
lightrays near $r=0$ is shown in Fig.~\ref{cap:build lines} (right). It is
clear from the figure that in a conformal diagram the timelike boundary must
be represented by a line with a slope between $45^{\circ}$ and $135^{\circ}$.
We shall call such lines \textsl{vertically-directed}. 

\begin{figure}
\begin{center}\psfrag{x}{$x$}\psfrag{x1}{$x_1$}\psfrag{A}{$A$}\psfrag{B}{$B$}\psfrag{C}{$C$}\psfrag{D}{$D$}

\psfrag{x2}{$x_2$}\psfrag{t}{$t$}

\psfrag{t0}{$t_0$}\psfrag{F}{$F$}

\psfrag{r=0}{$r=0$}\includegraphics[%
  width=3in]{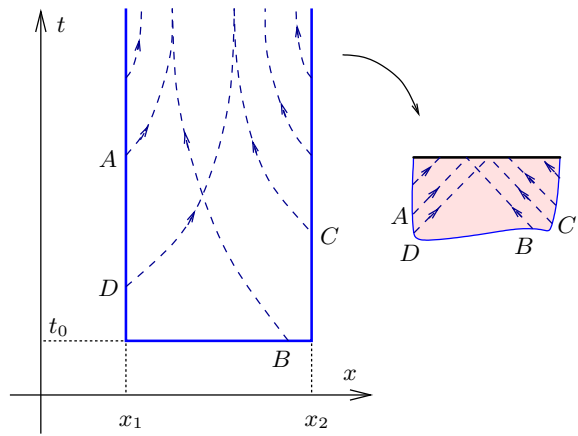}\end{center}

\caption{Construction of a conformal diagram for a part of de Sitter spacetime delimited
by thick lines (left). The upper boundary of the diagram (right) is a horizontal
line. \label{cap:de Sitter 1}}
\end{figure}

As an example of using timelike boundaries, let us consider the subdomain $\left(t_{0}<t<\infty,x_{1}<x<x_{2}\right)$
of a de Sitter spacetime with flat spatial sections, described by the metric\begin{equation}
g_{ab}dx^{a}dx^{b}=dt^{2}-e^{2Ht}dx^{2}.\label{eq:dS metric}\end{equation}
This subdomain can be visualized as the future of a selected initial comoving
region. The Cauchy line $t=t_{0}$ is connected to the two timelike boundary
lines, $x=x_{1,2}$. In the coordinate system~(\ref{eq:dS metric}), the null
geodesics are solutions of $dx/dt=\pm e^{-Ht}$ and it is easy to see that a
lightray emitted at $x=0$, $t=0$ only reaches the values $\left|x\right|<H^{-1}$
(the limit value is the de Sitter horizon). Null geodesics emitted from the
Cauchy surface and from the boundary lines are sketched in Fig.~\ref{cap:de Sitter 1}
where it is assumed that the comoving domain $x_{1}<x<x_{2}$ contains several
de Sitter horizons. A lightray emitted in the positive direction, such as the
ray $A$, intersects left-directed lightrays emitted from the point $B$ or
from nearer points but does not intersect lightrays emitted further away, such
as the ray $C$. We call the ray $B$ the \textsl{rightmost ray} intersecting
$A$. It is clear that the intersection of the corresponding lines $A$ and
$B$ in the conformal diagram must occur \emph{at the boundary} of the diagram,
otherwise there would exist further lines intersecting $A$ to the right of
$B$. Considering a ray $D$ to the right of $A$, we find that the rightmost
ray for $D$ is $C$. The intersection of $C$ and $D$ is thus another point
on the boundary of the conformal diagram. It follows that the boundary line
must have a slope between $-45^{\circ}$ and $45^{\circ}$; for simplicity,
we draw a straight horizontal line (Fig.~\ref{cap:de Sitter 1}, right). This
line represents the (timelike and null) infinite future.

A timelike boundary can be interpreted as the trajectory of an observer who
absorbs or emits lightrays and thus participates in the exploration of the causal
structure of the spacetime. The lightrays emitted by the boundary, together
with those emitted from the Cauchy surface, form the totality of all lightrays
that must be bijectively mapped into straight lines in the conformal diagram.
The role of timelike boundaries and Cauchy surfaces is to provide a physically
motivated boundary for the part of the spacetime we are interested in.

An artificial timelike boundary may also be introduced into the spacetime with
the purpose of simplifying the construction of the diagram. Below (Sec.~\ref{sub:Many-bubbles})
we shall show that conformal diagrams can be pasted together along a common
timelike boundary.

\subsection{General procedure\label{sub:General-procedure}}

We can now outline a general procedure for building a conformal diagram for
a given 1+1-dimensional spacetime using the method of lightrays. The procedure
can be applied not only to geodesically complete spacetimes but also to spacetimes
with explicitly specified boundaries.

One starts by considering the future-directed part of the spacetime and by choosing
a suitable Cauchy surface and, possibly, some timelike boundaries. These lines
are represented in the diagram by arbitrarily drawn horizontal and vertical
curves of finite extent. The endpoints of these curves correspond either to
the intersection points of Cauchy lines and timelike boundaries, or to imaginary
points at spacelike and timelike infinity. 

Note that Cauchy surfaces and timelike boundaries are the lines on which boundary
conditions for e.g.~a wave equation must be specified to obtain a unique solution
within a domain. One expects that any physically relevant spacetime should contain
a suitable set of Cauchy surfaces and timelike boundaries, if the classical
field theory is to have predictive power. Qualitative knowledge of the geometry
and topology of these boundaries is required for building a conformal diagram
using the method of lightrays.

After drawing the curves for the physical boundaries, it remains to determine
the shape of the infinite (timelike and null) boundaries of the diagram. To
this end, one can first consider \emph{right-directed} lightrays emitted from
various points on the Cauchy line and from the boundaries, including the limit
points at infinity. For each right-directed lightray $X$ there exists a certain
subset ${\cal S}_{X}$ of (left-directed) lightrays that intersect $X$. Since
the subset ${\cal S}_{X}$ has a finite extent, there exists a rightmost ray
$r_{X}\in{\cal S}_{X}$. (In the de Sitter example above, the rightmost ray
$r_{A}$ is the ray $B$ and the rightmost ray $r_{D}$ is $C$.) By definition
of the conformal diagram, the lightrays are straight lines limited only by the
boundary of the conformal diagram, therefore the intersection point of $X$
and $r_{X}$ must belong to the boundary. In this way we have established the
location of one point of the unknown boundary line, namely the endpoint of the
ray $X$.

\begin{figure}
\begin{center}\psfrag{A}{(a)}\psfrag{B}{(b)}\psfrag{C}{$C$}\psfrag{D}{$D$}

\psfrag{X}{$X$}\psfrag{rx}{$r_X$}

\psfrag{X1}{$X'$}\psfrag{rx1}{$r_{X'}$}\includegraphics[%
  width=3in]{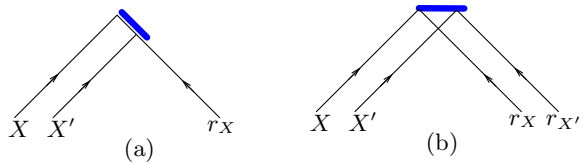}\end{center}

\caption{The local direction of the infinite boundary (thick line) is found by determining
the rightmost rays $r_{X}$, $r_{X'}$ for nearby rays $X,X'$. The direction
is at $45^{\circ}$ angle (a) when $r_{X}=r_{X'}$ and horizontal (b) when $r_{X}\neq r_{X'}$.\label{cap:general procedure}}
\end{figure}

To determine the local direction of the boundary line at that point, we use
the following argument. For each right-directed lightray $X$, we can consider
a right-directed ray $X'$ infinitesimally close and to the right of $X$ (if
no ray $X'$ can be found to the right of $X$, it means that $X$ itself belongs
to the boundary of the diagram). Then there are two possibilities (see Fig.~\ref{cap:general procedure}a,b):
either the rightmost ray $r_{X}$ is also the rightmost ray $r_{X'}$ for $X'$,
or the ray $r_{X'}$ is located to the right of $r_{X}$. In the first case,
the boundary line has a $45^{\circ}$ slope and locally coincides with $r_{X}$,
while in the second case the boundary line is horizontally-directed. Thus we
can draw a right-directed fragment of the infinite boundary that limits the
ray $X$. We then continue by moving further to the right and consider the endpoint
of the ray $X'$, etc.

The same procedure is then repeated for left-directed lightrays, until one finishes
drawing all unknown lines in the future-directed part of the diagram. In this
way the infinite boundary of the conformal diagram is constructed as the locus
of {}``last intersections'' of lightrays. Finally one applies the same considerations
to past-directed lightrays and so completes the diagram.

It follows that the infinite boundary of a conformal diagram can always be drawn
as a sequence of either straight line segments directed at $45^{\circ}$ angles,
or horizontally- and vertically-directed curves. In our convention, Cauchy surfaces
and artificial timelike boundaries are drawn as curved lines and infinite boundaries
as straight lines (when possible).

\subsection{Further examples}

\subsubsection*{Collapsing star}

The method of lightrays does not require explicit formulae for the spacetime
metric if the qualitative behavior of lightrays is known. As another example
of using the lightray method, let us consider an asymptotically flat spacetime
with a star collapsing to a black hole (BH).

\begin{figure}
\begin{center}\psfrag{A}{$A$}\psfrag{B}{$B$}\psfrag{C}{$C$}\psfrag{D}{$D$}

\psfrag{E}{$E$}\psfrag{F}{$F$}

\psfrag{E1}{$E'$}\psfrag{F1}{$F'$}\psfrag{F2}{$F''$}

\psfrag{L}[Bl][Bl][1][12]{Cauchy line}

\psfrag{r=0}{$r=0$}\includegraphics[%
  width=2in]{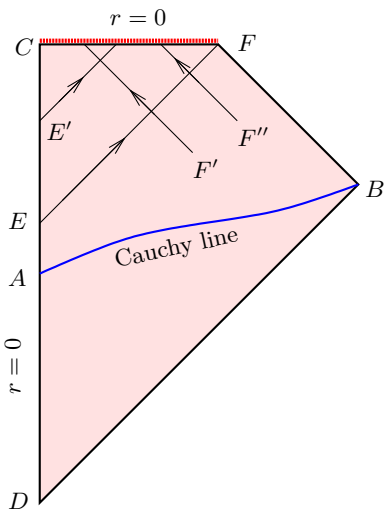}\end{center}

\caption{Construction of the conformal diagram for the spacetime of a collapsing star.
The thick dotted line $CF$ is a Schwarzschild singularity. \label{cap:build star}}
\end{figure}

To reduce the spacetime to 1+1 dimensions, we assume spherical symmetry and
consider only the $\left(r,t\right)$ plane, where $0<r<+\infty$ and the line
$r=0$, the center of the star, is an artificial boundary. As before, we start
with the future-directed part of the diagram. We must first choose a Cauchy
surface; a suitable Cauchy surface is the line $t=t_{0}$ where $t_{0}$ is
a time chosen before the collapse of the star. We represent the Cauchy surface
by the curve $AB$ in the diagram (Fig.~\ref{cap:build star}). The point $A$
corresponds to ($r=0$, $t=t_{0}$), while $B$ is a spatial infinity $\left(r=\infty,t=t_{0}\right)$.
The artificial boundary $r=0$ is represented by the vertical line $AC$.

To investigate the shape of the future part of the diagram, we need to analyze
the intersections of lightrays emitted from faraway points of the Cauchy surface.
We know from qualitative considerations of the black hole formation that lightrays
can escape from the star interior only until the appearance of the BH horizon.
Shortly thereafter the star center becomes a singularity that cannot emit any
lightrays. Hence, among all the rays emitted from the star center at various
times, there exists a {}``last ray'' not captured by the BH, while rays emitted
later are captured. We arbitrarily choose a point $E$ on the boundary line
$r=0$ to represent the emission of this {}``last ray.'' Any lightray emitted
from $r=0$ before $E$ will propagate away from the BH and so will intersect
all left-directed lightrays emitted from arbitrarily remote points of the Cauchy
line $AB$. Since all the intersection points are outside of the BH horizon,
it follows that the conformal diagram contains the polygon $AEFB$ which represents
the spacetime outside the BH. The line $FB$ is the infinite null boundary of
the diagram, while the line $EF$ is the BH horizon.

It is clear that the point $C$ is the last point from which lightrays can be
emitted from the star center and thus $C$ is the beginning of the BH singularity.
It remains to determine the shape of the diagram between the points $C$ and
$F$. We know that a lightray emitted from the center after $E$ will be recaptured
by the BH singularity and thus will not intersect with lightrays entering the
BH horizon sufficiently late. For instance, the lightray emitted at $E'$ will
intersect with the lightray $F'$ but not with a later ray $F''$, as shown
in Fig.~\ref{cap:build star}. Therefore the diagram boundary line connecting
$C$ and $F$ is locally horizontally-directed. This line consists of final
intersection points of lightrays emitted from the star center and those entering
the BH horizon from outside. These final intersection points are located at
the BH singularity which is therefore represented by the entire line $CF$. 

The past-directed part of the conformal diagram is easy to construct. Since
any two past-directed lightrays intersect, the past-directed part is similar
to that for the Minkowski spacetime with a boundary at $r=0$, namely the triangular
domain $ABD$. Thus the diagram in Fig.~\ref{cap:build star} is complete.

\subsubsection*{Future part of de Sitter spacetime}

In Fig.~\ref{cap:de Sitter 1} we have drawn a conformal diagram for the subdomain
of a de Sitter spacetime delimited by two timelike boundaries. We shall now
construct the diagram for the future part of a de Sitter spacetime with spatially
unlimited sections. (Note that the past half of the de Sitter spacetime is not
covered by the flat coordinates because of incompleteness of past-directed geodesics.
In this paper we shall not use the complete de Sitter spacetime but only the
future of an arbitrarily chosen, unbounded, spacelike Cauchy hypersurface.)

\begin{figure}
\begin{center}\psfrag{A}{$A$}\psfrag{B}{$B$}\psfrag{t}{$t=t_0$}\includegraphics[%
  width=3in]{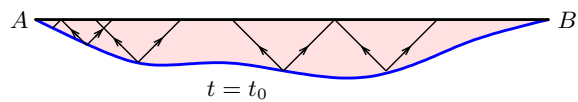}\end{center}

\caption{A conformal diagram for the future part of the de Sitter spacetime with flat
spatial sections. The curved line represents the spatially infinite Cauchy surface
$t=t_{0}$. The local behavior of lightrays is the same as in Fig.~\ref{cap:de Sitter 1}.
\label{cap:de Sitter3}}
\end{figure}

The Cauchy surface $t=t_{0}$, $-\infty<x<\infty$ is drawn as a finite horizontal
curve in the conformal diagram (the curved line $AB$ in Fig.~\ref{cap:de Sitter3}).
The points $A,B$ in the diagram represent a spacelike infinity in the two directions
and do not correspond to any points in the physical spacetime. It remains to
establish the shape of the infinite future boundary which must be a line connecting
the points $A,B$ to the future of the Cauchy surface. We already know from
the construction of the diagram in Fig.~\ref{cap:de Sitter 1} that this future
boundary is locally horizontal. Since the behavior of lightrays emitted from
all points of the Cauchy surface is the same, the future boundary may be represented
by a horizontal straight line connecting the points $A,B$. (To keep the convention
of having straight infinite boundaries, we have drawn the Cauchy surface $t=t_{0}$
as a curve extending downward from the straight line $AB$.)

Note that the same diagram also represents an inhomogeneous spacetime consisting
of de Sitter-like regions with different values of the Hubble constant $H$.
This is so because a difference in the local values of $H$ does not change
the qualitative behavior of lightrays at infinity: the rays will intersect only
if emitted from sufficiently near points. The infinite future boundary remains
a horizontally-directed line as long as $H\neq0$ everywhere.

The construction of conformal diagrams for the following spacetimes is left
as an exercise for the reader.

\begin{figure}
\begin{center}\psfrag{A}{$A$}\psfrag{B}{$B$}\psfrag{x1}[Bl][Bl][1][90]{$x=x_1$} \psfrag{x2}[Bl][Bl][1][90]{$x=x_2$}\includegraphics[%
  width=3in]{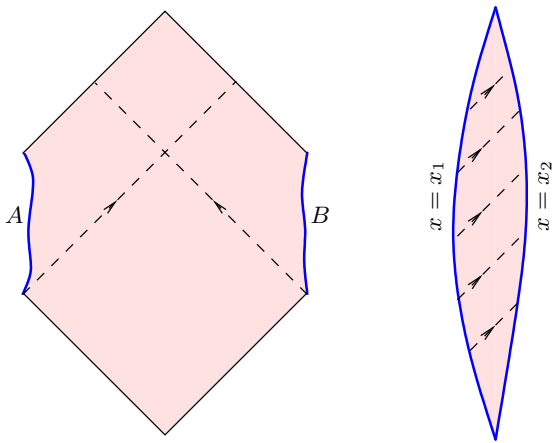}\end{center}

\caption{Left: The domain of Minkowski spacetime between two causally separated observers
$A,B$ moving with a constant proper acceleration in opposite directions. The
dashed lines show that the two observers cannot see each other. Right: A closed
Minkowski universe. The thick lines represent the identified boundaries $x=x_{1,2}$.
A lightray (dashed line) crosses the line $x=x_{1,2}$ infinitely many times.
\label{cap:Mink3}}
\end{figure}

\begin{figure}
\begin{center}\psfrag{A}{$A$}\psfrag{B}{$B$}\psfrag{T}{$T$}\includegraphics[%
  width=2in]{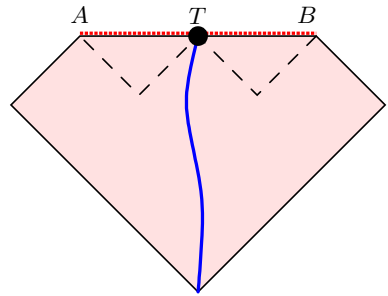}\end{center}

\caption{A spacetime with two collapsing stars. The point $T$ is a future timelike
infinity reached by an observer (thick curve) remaining between the two black
holes. The lines $AT$ and $TB$ represent BH singularities and the dashed lines
are the BH horizons. \label{cap:2BH}}
\end{figure}

\begin{figure}
\begin{center}\psfrag{A}{$A$}\psfrag{B}{$B$}\psfrag{T}{$T$}\includegraphics[%
  width=2in]{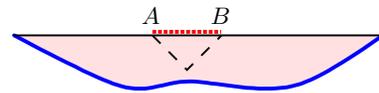}\end{center}

\caption{The future part of a de Sitter spacetime with a collapsing star. The line $AB$
represents the BH singularity; the dashed line is the BH horizon. \label{cap:dsBH}}
\end{figure}

\begin{figure}
\begin{center}\includegraphics[%
  width=3.2in]{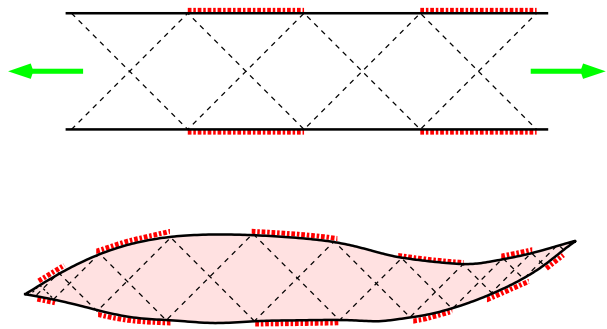}\end{center}

\caption{Diagrams for a maximally extended Schwarzschild-de Sitter spacetime. The conventional,
unbounded diagram (top) and an equivalent diagram having a finite extent (bottom)
are related by a conformal transformation. The thick dotted lines represent
Schwarzschild singularities. Both diagrams contain infinitely many Schwarzschild
and de Sitter regions. \label{cap:SdS}}
\end{figure}

\begin{itemize}
\item A subdomain of the Minkowski spacetime between two causally separated observers
moving with a constant proper acceleration in opposite directions (Fig.~\ref{cap:Mink3},
left).
\item A flat closed universe: the subdomain ($-\infty<t<\infty$, $x_{1}<x<x_{2}$)
of a Minkowski spacetime with the lines $x=x_{1}$ and $x=x_{2}$ identified
(Fig.~\ref{cap:Mink3}, right). 
\item An asymptotically flat spacetime with two stars collapsing into two black holes;
the line of sight crosses the two star centers (Fig.~\ref{cap:2BH}).
\item The future part of a de Sitter spacetime with a star collapsing into a black
hole (Fig.~\ref{cap:dsBH}).
\item A maximally extended Schwarzschild-de Sitter spacetime (Fig.~\ref{cap:SdS}).
It is interesting to note that this diagram is usually drawn with all Schwarzschild
and de Sitter regions having the same size, which makes the figure unbounded
(Fig.~\ref{cap:SdS}, top) despite the intention to represent the spacetime
by a \emph{finite} figure in the fiducial Minkowski plane. To adhere to the
definition of a conformal diagram as a bounded figure, one can use a suitable
conformal transformation reducing the diagram to a finite size (e.g.~Fig.~\ref{cap:SdS},
bottom).
\end{itemize}

\section{Diagrams for eternally inflating spacetimes}

\begin{figure}
\begin{center}\psfrag{t}{$t$}\psfrag{x1}{$x_1$}\psfrag{x2}{$x_2$}

\psfrag{x}{$x$}\includegraphics[%
  width=1.5in]{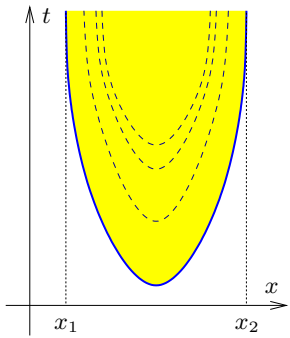}\end{center}

\caption{A thermalized bubble in a de Sitter spacetime with flat spatial sections. The
dashed lines are equal-temperature surfaces in the thermalized region (shaded).
\label{cap:bubble 1}}
\end{figure}

I now apply the method of lightrays to spacetimes resulting from generic models
of eternal inflation. In these models, inflation ends in a particular region
of spacetime but continues in nearby regions. The process is analogous to the
nucleation of non-inflating bubbles in a de Sitter spacetime. More precisely,
the bubble interior inflates at a progressively slower rate until the inflation
stops and the interior reheats. In the meantime, the bubble continues to expand,
annexing new regions that stop inflating somewhat later. The bubble wall moves
with a constant acceleration and its worldline quickly becomes almost lightlike.
As was noted in Ref.~\cite{CdL80}, the interior of the bubble appears to be
an infinite FRW universe to an interior observer, even though the bubble occupies
a finite interval $x_{1}<x<x_{2}$ of the comoving volume in the background
de Sitter spacetime (see Fig.~\ref{cap:bubble 1}). In particular, the hypersurfaces
of equal temperature (identified as the equal-time surfaces by interior observers)
have an infinite proper volume.

In an eternally inflating spacetime, all Hubble-sized regions are causally disconnected
and thus the nucleation of thermalized bubbles is approximately a Poisson process
with a fixed nucleation rate per proper 4-volume. We first construct a conformal
diagram for an inflating spacetime with one thermalized bubble and then generalize
to a random arrangement of such bubbles. We shall then consider models where
the thermalized bubbles eventually become dominated by dark energy and again
become patches of the de Sitter spacetime.

\subsection{Spacetime with one bubble}

As we have already noted, inflating spacetimes with an inhomogeneous expansion
rate are represented by the same conformal diagram as a homogeneous de Sitter
spacetime. On the other hand, regions where inflation is finished will have
the same causal structure as a flat spacetime. Therefore we can ignore the details
of reheating and concentrate on building a conformal diagram for a pure de Sitter
spacetime ($H=H_{0}$) with a nucleating Minkowski bubble ($H=0$), which is
a considerably simpler situation. Since the thickness of bubble walls plays
no role in the shape of the resulting conformal diagrams, we shall always treat
bubbles in the thin-wall approximation.

We start by choosing a Cauchy surface for the future part of the diagram. An
appropriate Cauchy surface would be $t=t_{0}$ where $t_{0}$ is a time moment
before the bubble formation. This surface is represented by a spacelike curve
in the diagram. We now need to analyze trajectories of lightrays emitted from
various points of the Cauchy surface and to determine whether any given pair
of lightrays will eventually intersect. As we have seen in Sec.~\ref{sub:General-procedure},
the future infinite boundary of the conformal diagram can be constructed as
the locus of {}``last intersections'' of lightrays.

\begin{figure}
\begin{center}\psfrag{t}{$t$}\psfrag{x1}{$x_1$}\psfrag{x2}{$x_2$}

\psfrag{x}{$x$}\psfrag{?}{?}\psfrag{A}{$A$}\psfrag{B}{$B$}\includegraphics[%
  width=3in]{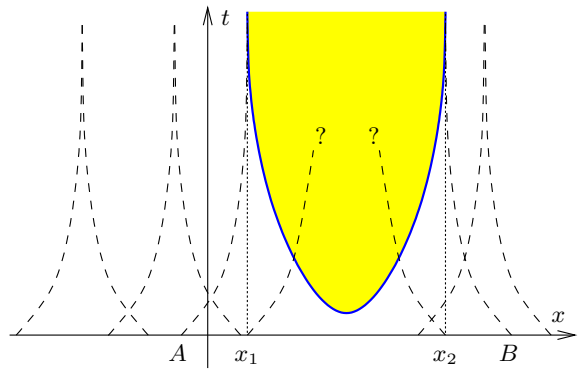}\end{center}

\caption{Null geodesics (dashed lines) in a de Sitter spacetime with a bubble (shaded).
The last lightrays not entering the bubble are labeled $A$ and $B$. Question
marks indicate that the behavior of lightrays in the bubble interior is yet
to be investigated. \label{cap:dS4}}
\end{figure}

\begin{figure}
\begin{center}\psfrag{T}{$T$}\psfrag{C}{$C$}\psfrag{D}{$D$}

\psfrag{A}{$A$}\psfrag{B}{$B$}\includegraphics[%
  width=2.5in]{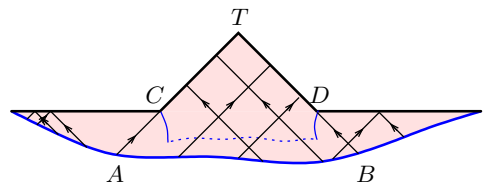}\end{center}

\caption{A conformal diagram for the future part of de Sitter spacetime with a dust-dominated
bubble. The bubble walls are the vertical curves ending at $C$ and $D$; the
(spacelike) dotted line symbolizes the nucleation event.\label{cap:dS5}}
\end{figure}

Typical lightray trajectories are sketched in Fig.~\ref{cap:dS4}. As in the
pure de Sitter spacetime, lightrays not entering the bubble intersect only if
emitted from sufficiently near points. Therefore the rightmost and leftmost
lightrays not yet entering the bubble (labeled $A,B$ in Fig.~\ref{cap:dS4})
delimit the parts of the conformal diagram that coincide with a pure de Sitter
diagram (Fig.~\ref{cap:de Sitter3}). Thus the future boundary of our diagram
contains two horizontal lines as shown in Fig.~\ref{cap:dS5}. The points $C,D$
correspond to the asymptotic comoving positions of the bubble walls ($x=x_{1,2}$).

It remains to determine the behavior of lightrays that do enter the bubble.
We shall shortly demonstrate that any two such lightrays necessarily intersect
if they enter the bubble from opposite sides. It will then follow that the ray
$B$ in Fig.~\ref{cap:dS4} is the rightmost ray for all rays entering the
bubble from the left, and likewise the ray $A$ is the leftmost ray for all
rays entering from the right. Considerations similar to those used in the construction
of the Minkowski diagram (Sec.~\ref{sub:Minkowski-spacetime}) will then show
that the infinite boundary of the diagram between the points $C$ and $D$ consists
of two null lines $CT$ and $TD$, while the point $T$ represents a timelike
infinity for observers inside the bubble (see Fig.~\ref{cap:dS5}). 

We shall first give a qualitative argument showing that any two lightrays entering
the bubble will eventually intersect. The bubble interior is a certain subdomain
of the 1+1-dimensional Minkowski spacetime. Any two lightrays entering the bubble
from opposite sides will enter the flat subdomain at two spacelike-separated
points and will be propagating towards each other. So these rays will eventually
intersect somewhere inside the bubble. 

\begin{figure}
\begin{center}\psfrag{t}{$\tau$}\psfrag{x}{$\zeta$}\includegraphics[%
  width=2in]{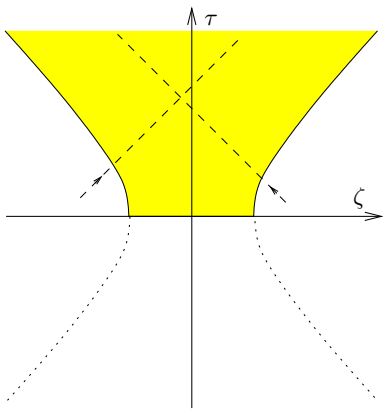}\end{center}

\caption{The bubble interior (shaded) viewed in the flat coordinates $\left(\tau,\zeta\right)$.
The dotted lines show the continuation of the bubble wall to times before the
bubble nucleation ($\tau=0$). The dashed lines are lightrays entering the bubble
from opposite sides; such lightrays always intersect. \label{cap:bubble interior}}
\end{figure}

To present a more explicit geometric argument, we need to consider the geometry
of the bubble interior in some detail. A 1+1-dimensional de Sitter spacetime
can be embedded in a flat three-dimensional spacetime with coordinates $\left(\tau,w,\zeta\right)$
and the metric $ds^{2}=d\tau^{2}-dw^{2}-d\zeta^{2}$ as a hyperboloid \begin{equation}
\zeta^{2}+w^{2}-\tau^{2}=H_{0}^{-2}\label{eq:dS hyp}\end{equation}
 (we follow the notation of Ref.~\cite{GarVil98}). The geometry of the one-bubble
spacetime can be visualized as the union of a part of the de Sitter hyperboloid
and a part of a hyperplane $w=w_{0}$ intersecting the hyperboloid, where $w_{0}^{2}<H_{0}^{-2}$.
The intersection line specified by the equations\begin{equation}
\zeta^{2}-\tau^{2}=H_{0}^{-2}-w_{0}^{2},\quad w=w_{0},\end{equation}
cuts out the subdomain $\zeta^{2}-\tau^{2}<H_{0}^{-2}-w_{0}^{2}$ of the Minkowski
hyperplane $\left(\tau,\zeta,w=w_{0}\right)$. This subdomain therefore contains
the interior of the bubble (Fig.~\ref{cap:bubble interior}). It is then evident
that any two opposite lightrays entering the bubble will certainly intersect
within its interior. This argument concludes the construction of the conformal
diagram shown in Fig.~\ref{cap:dS5}.

\subsection{Dark energy-dominated bubble}

Let us now consider a cosmological scenario where the thermalized bubble interior
eventually becomes dominated by dark energy. If the vacuum energy density is
positive then the bubble interior becomes a patch of a de Sitter spacetime.
The case of negative dark energy will be considered in the next subsection.

\begin{figure}
\begin{center}\psfrag{A}{$A$}\psfrag{B}{$B$}\psfrag{t}{$\tau$}\psfrag{x}{$\zeta$}\includegraphics[%
  width=2in]{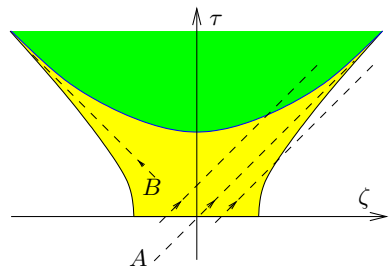}\end{center}

\caption{The bubble interior as in Fig.~\ref{cap:bubble interior} but with a dark
energy-dominated subdomain (darker shade). The rays labeled $A,B$ asymptotically
coincide with the bubble walls. \label{cap:bubble interior 2}}
\end{figure}

\begin{figure}
\begin{center}\psfrag{E}{$E$}\psfrag{C}{$C$}\psfrag{D}{$D$}

\psfrag{A}{$A$}\psfrag{B}{$B$}\psfrag{F}{$F$}\includegraphics[%
  width=2.5in]{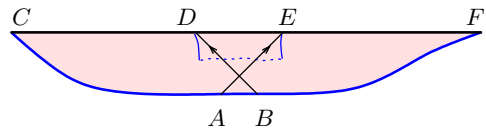}\end{center}

\caption{A conformal diagram for the future part of de Sitter spacetime with a dark-energy
dominated bubble interior. The bubble walls are shown as in Fig.~\ref{cap:dS5}.
The points $D,E$ represent the asymptotic position of the bubble walls. \label{cap:dS6}}
\end{figure}

The epoch of dark energy domination begins at an equal-temperature hypersurface
at a certain cosmological time $t_{DE}$ after the bubble nucleation. Due to
the $SO(3,1)$ symmetry of the bubble, the dark energy-dominated domain is specified
in the flat coordinates $\left(\tau,\zeta\right)$ by the inequality\begin{equation}
\tau^{2}-\zeta^{2}>t_{DE}^{2},\end{equation}
and the boundary $t^{2}-\zeta^{2}=t_{DE}^{2}$ approaches the lightcone $\tau=\pm\zeta$
for large $\tau$. It then follows from Fig.~\ref{cap:bubble interior 2} that
the lightray labeled $A$ separates the right-directed rays exiting the bubble
from rays entering and remaining within the dark energy-dominated subdomain.
Since the lightray $A$ asymptotically coincides with the bubble wall, the endpoint
$E$ of that ray in a conformal diagram (Fig.~\ref{cap:dS6}) divides the future
infinite boundary $CF$ of the diagram into the parts representing the future
of the exterior de Sitter spacetime ($EF$) and the part corresponding to the
interior de Sitter subdomain. Similarly, the ray $B$ divides left-directed
rays exiting the bubble from those entering the de Sitter subdomain. The infinite
future boundary of the de Sitter subdomain is the same as that in Fig.~\ref{cap:de Sitter3}.
Therefore a conformal diagram for the spacetime with a dark-energy dominated
bubble can be drawn as in Fig.~\ref{cap:dS6} and is the same as the diagram
for the future part of a pure de Sitter spacetime. The points $D,E$ are not
special with respect to the causal structure and are marked only to specify
the asymptotic locations of the bubble walls.

\subsection{Bubble dominated by negative dark energy}

The last possibility to consider is a bubble that becomes dominated by a negative
dark energy ({}``anti-de Sitter bubble''). This leads to a collapse of the
bubble interior into a singularity~\cite{CdL80,AC85}. Again assuming the $SO(3,1)$
symmetry, we find that the singularity occurs along a spacelike hypersurface
$t^{2}-\zeta^{2}=t_{s}^{2}$, where $t_{s}$ is the cosmological time of collapse.
The geometry of the singular surface is quite similar to that of the surface
of dark-energy domination, so the singularity may be represented in a conformal
diagram by a straight horizontal line connecting the asymptotic positions of
the bubble walls. Therefore Fig.~\ref{cap:dsBH} may be reused as a conformal
diagram for a bubble dominated by negative dark energy if we reinterpret the
line $AB$ as a cosmological singularity rather than a Schwarzschild singularity.

\subsection{Many bubbles\label{sub:Many-bubbles}}

Before turning to many-bubble spacetimes, we note that the diagram in Fig.~\ref{cap:dS5}
can be obtained by pasting together the diagram in Fig.~\ref{cap:de Sitter 1}
(right) and the top half of the diagram in Fig.~\ref{cap:Mink3} (left). The
pasting is performed along the timelike bubble walls treated as artificial boundaries.

Let us briefly consider the pasting of conformal diagrams in general. When two
spacetime domains are separated by a timelike worldline, the corresponding conformal
diagrams can be pasted together along the boundary line. To verify this almost
obvious statement more formally, we begin by drawing the timelike boundary as
a vertically-directed line in the diagram. The shape of the conformal diagram
to the right of the boundary is determined solely by the intersections of lightrays
within the right half of the spacetime. Hence, the conformal diagram for the
right half of the spacetime can be pasted to the right of the timelike boundary
line. The same holds for the left half of the spacetime. This justifies pasting
of diagrams along a common timelike boundary.

\begin{figure}
\begin{center}\includegraphics[%
  width=2.5in]{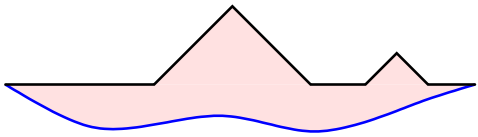}\end{center}

\caption{A conformal diagram for the future part of de Sitter spacetime with two dust-dominated
bubbles. \label{cap:dS7}}
\end{figure}

It follows that a conformal diagram for a spacetime with \emph{two} nucleating
bubbles is a simple concatenation of two diagrams drawn for single-bubble spacetimes.
For instance, a diagram for two (nonintersecting) dust-dominated bubbles is
shown in Fig.~\ref{cap:dS7}. It features two {}``roofs'' representing null
and timelike infinities of the two bubble interiors.

Now we consider an eternally inflating spacetime that contains infinitely many
thermalized bubbles at random locations and times. In the comoving coordinates,
the walls of each bubble asymptotically approach lines $x=\textrm{const}$ at
late times. We may collect all such lines $x=x_{i}$, $i=1,2,...$, into a set
$E\equiv\left\{ x_{i}\right\} $. This set consists of all comoving trajectories
that never enter any thermalized regions. The set $E$ was called the {}``eternal
set'' in Ref.~\cite{Win02} where it was shown that $E$ is topologically
closed, its comoving measure is zero, and any neighborhood of any point $x_{i}\in E$
contains infinitely many other points from $E$ (the fractal property). The
asymptotic late-time bubble interiors are comoving intervals of $x$ that do
not contain any points from $E$. Infinitely many such intervals cover the entire
length of the $x$ axis.

\begin{figure}
\begin{center}\psfrag{0}{0}\psfrag{1}{1}\includegraphics[%
  width=3in]{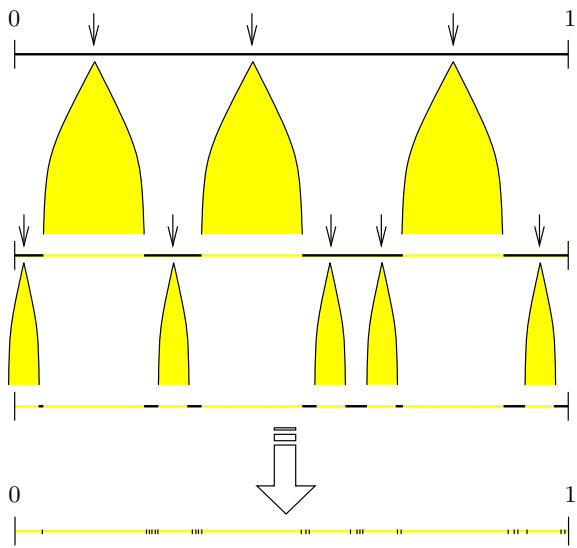}\end{center}

\caption{Construction of a random Cantor set. Arrows show the bubble nucleation sites
randomly chosen at each timestep. The (comoving) bubble size decreases geometrically
with time and the number of nucleation sites increases. After infinitely many
steps, there remains a fractal set of points (sketched below). \label{cap:setE}}
\end{figure}

One can visualize the set $E$ as a random Cantor set (see e.g.~\cite{Mandelbrot},
chapters 8, 23, and 31). A procedure to construct a suitable Cantor set consists
of simulating the stochastic process of bubble nucleation (Fig.~\ref{cap:setE}).
It is known that bubble intersections are rare in realistic models~\cite{GW83},
so we shall for simplicity assume that each bubble expands to the Hubble horizon
size in the comoving coordinates. For the purposes of simulation, we can start
with an interval $\left[0,1\right]$ of comoving space and divide the time into
discrete steps. At each timestep, we randomly choose some bubble nucleation
sites using a Poisson process with a fixed number of bubbles per horizon length.
At the next step the bubbles grow to horizon size, so the corresponding segments
are removed from the interval and new nucleation sites are chosen. The horizon
size decreases geometrically with time and thus the number of removed intervals
grows and their length decreases. The set $E$ consists of points that have
not been removed after infinitely many steps (computer time permitting). 

\begin{figure}
\begin{center}\includegraphics[%
  width=3in]{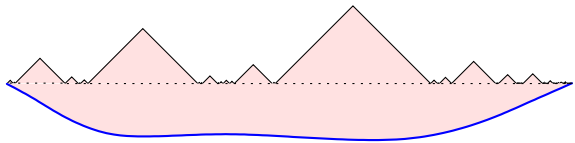}\end{center}

\caption{A conformal diagram for the future part of an eternally inflating spacetime
with a fractal arrangement of dust-dominated bubbles. (A computer simulation
was used to position the bubbles.) \label{cap:dS8}}
\end{figure}

In the conformal diagram, the infinite future of each bubble interior is represented
by a {}``roof'' if the bubble is dust-dominated and by a horizontal line if
the bubble is dark energy-dominated. Since the latter case results in a rather
uninteresting diagram (the same as in Fig.~\ref{cap:dS6} but with infinitely
many bubble walls), we shall confine our attention to dust-dominated bubbles.
A conformal diagram for that case is shown in Fig.~\ref{cap:dS8} and can be
constructed as follows. First we imagine a pure de Sitter region without bubbles
and draw its infinite future boundary (dashed line in Fig.~\ref{cap:dS8}).
Then the asymptotic positions of bubble walls are marked on that line as the
points of a random Cantor set $E$. Finally, each interval not containing any
points from $E$ is raised to a {}``roof'' representing a null and a timelike
infinity of the bubble interior. 

It follows from the fractal property of the set $E$ that there exist infinitely
many arbitrarily small roofs between any two roofs in Fig.~\ref{cap:dS8}.
Thus a conformal diagram for an eternally inflating spacetime must contain a
fractal arrangement of infinitely many lines.

I conclude this section by a comment on the relevance of 1+1-dimensional diagrams
to 3+1-dimensional universes. As I have noted above, the reduction to a 1+1-dimensional
slice is useful only if all null geodesics within the slice are also geodesics
in the full spacetime. This is the case for a spherically symmetric spacetime
reduced to the $\left(t,r\right)$ slice, but generally not for a $\left(t,x\right)$
slice of a $\left(t,x,y,z\right)$ spacetime with a random distribution of bubbles,
even if each bubble were spherically symmetric. However, we are presently considering
an inflationary spacetime with a geometry that is approximately de Sitter in
large (super-Hubble) domains. Then one can define the local expansion rate $H(t,x,y,z)$
which is a slowly-changing function in both space and time. In this case the
null lines within an arbitrary 1+1-dimensional slice, such as the slice $\left(t,x\right)$,
are approximately geodesic lines in the full spacetime. Therefore the diagram
in Fig.~\ref{cap:dS8} can be used as a qualitative visualization of the causal
structure of such a spacetime along a randomly chosen line of sight.

\section{Diagrams for recycling universes}

In quantum field theory there exist several possibilities for a transition,
via bubble nucleation, from one de Sitter (dS) state to another dS state having
a different vacuum energy~\cite{CdL80,LeeWei87,GarMeg04}. The model of a recycling
universe~\cite{GarVil98} involves such {}``false-vacuum bubbles'' randomly
nucleating in regions of true vacuum. Within the nucleated dS bubbles, inflation
begins anew ({}``recycling'') and eventually creates new thermalized regions
of true vacuum. The process of recycling repeats \emph{ad infinitum}. Note that
the geometry resulting from the nucleation of false-vacuum bubbles is qualitatively
different from that resulting from a cosmological dark-energy domination, because
the latter occurs everywhere in the bubble interior, while nucleated bubbles
typically never merge to fill the entire spatial volume~\cite{GW83}.

The model of the string theory landscape~\cite{landscape} is a general cosmological
scenario where locally de Sitter-like regions can randomly nucleate asymptotically
flat ({}``Minkowski''), de Sitter, or {}``anti-de Sitter'' (AdS) bubbles.
The dS bubbles can nucleate further bubbles in their interior and thus give
rise to the recycling process. On the other hand, AdS regions (more precisely,
bubbles that are eventually dominated by a negative dark energy) will collapse
to a singularity. So we assume that no further bubbles can be nucleated within
AdS or Minkowski bubbles.

In this last section we consider cosmological scenarios involving the phenomenon
of recycling and draw the corresponding conformal diagrams.

We begin by building a conformal diagram for a flat spacetime with one nucleated
dS bubble (although this process does not happen spontaneously, the diagram
is still useful). It is convenient to choose a spherically symmetric slice $\left(t,r\right)$
with an artificial boundary $r=0$, and a Cauchy surface preceding the bubble
nucleation. Then the past-directed part of the diagram is the same as for the
Minkowski spacetime, so we concentrate on the future part.

\begin{figure}
\begin{center}\psfrag{A}{$A$}\psfrag{B}{$B$}\psfrag{C}{$C$}\psfrag{D}{$D$}

\psfrag{E}{$E$}\psfrag{F}{$F$}

\psfrag{E1}{$E'$}\psfrag{F1}{$F'$}\psfrag{F2}{$F''$}

\psfrag{r=0}{$r=0$}\includegraphics[%
  width=2in]{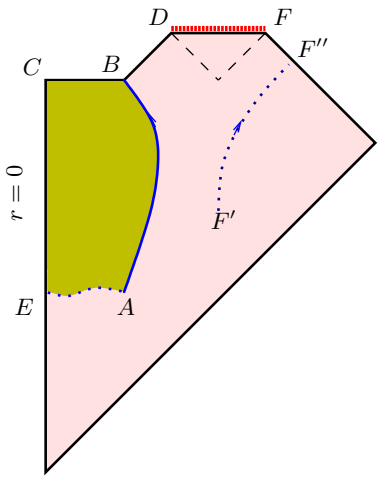}\end{center}

\caption{A flat spacetime with a nucleated de Sitter bubble. The bubble appears as a
black hole from outside (cf.~Fig.~\ref{cap:build star}) but contains a de
Sitter patch in its interior (shaded). The line $AB$ is the bubble wall, $EA$
is the nucleation surface, $DF$ is the BH singularity, $BC$ is the de Sitter
asymptotic future, and the dashed line is the BH horizon. The line $F^{\prime}F^{\prime\prime}$
is a trajectory accelerating away from the BH. \label{cap:star3}}
\end{figure}

The geometry of spherically symmetric false-vacuum bubbles embedded in Minkowski
regions is lucidly presented in Ref.~\cite{BGG87} and I shall merely outline
the results here. The bubble wall is moving with a constant acceleration towards
the region of higher expansion rate $H$. At a certain time, the bubble wall
implodes upon itself and disappears from view under a Schwarzschild horizon.
From that time onwards, the bubble appears to be a black hole when viewed from
outside. The corresponding part of the diagram is similar to that in Fig.~\ref{cap:build star},
and thus we can start drawing Fig.~\ref{cap:star3}. However, as explained
in Ref.~\cite{BGG87} and illustrated in Figs.~12-13 therein, the part of
the spacetime not accessible to exterior observers has a more complicated structure
than a usual BH interior. In fact this part of the spacetime contains a spatially
closed, causally disconnected universe which I shall call the {}``interior
universe'' for brevity. The interior universe contains subdomains of Schwarzschild
and of de Sitter (dS) spacetime. The bubble wall does not collapse onto the
BH singularity but passes into the interior universe where it continues to expand
with a constant acceleration, away from the inner Schwarzschild horizon. Beyond
the bubble wall there is an expanding subdomain of a de Sitter spacetime.

The bubble wall is a timelike boundary that allows us to combine the partial
de Sitter diagram for the bubble interior (Fig.~\ref{cap:de Sitter 1}, right)
with a diagram for the rest of the interior universe. To obtain the latter,
we forget about the bubble for a moment and consider a trajectory of an observer
uniformly accelerating away from a Schwarzschild horizon. This is the line $F^{\prime}F^{\prime\prime}$
in Fig.~\ref{cap:star3} where we have used the exterior Schwarzschild horizon
as an illustration. The range $FF^{\prime\prime}$ of the boundary line is the
null future for rays that escape from the BH but cannot overtake the accelerated
observer. A similar configuration occurs in the interior universe where the
bubble wall accelerates away from the inner BH horizon. Thus in the interior
universe there must exist lightrays escaping the inner BH horizon but never
overtaking the bubble wall. The null future of such lightrays is a line at $45^{\circ}$
angle (the line $BD$ in Fig.~\ref{cap:star3}) connecting the BH singularity
($DF$) with the asymptotic future of the bubble wall (point $B$).

This argument concludes the justification of the diagram in Fig.~\ref{cap:star3}
which is equivalent to the appropriate part of Fig.~15 in Ref.~\cite{BGG87}.

More realistically, we can consider high-$H$ bubbles nucleating within low-$H$
de Sitter regions. There are two qualitatively different cases: the nucleated
bubble may have either subhorizon or superhorizon size in the background dS
spacetime. In the present paper I shall not be concerned with the relative likelihood
of these processes but simply consider both cases. Since there is an unbounded
4-volume in which to nucleate bubbles, all possible nucleation processes will
occur infinitely many times, no matter how small the nucleation rate.

\begin{figure}
\begin{center}\psfrag{A}{$A$}\psfrag{B}{$B$}\psfrag{C}{$C$}\psfrag{D}{$D$}

\psfrag{E}{$E$}\psfrag{F}{$F$}

\psfrag{E1}{$E'$}\psfrag{F1}{$F'$}\psfrag{F2}{$F''$}

\psfrag{r=0}{$r=0$}\includegraphics[%
  width=2in]{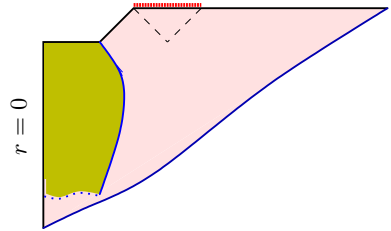}\end{center}

\caption{A future part of the dS spacetime with a dS bubble having a higher expansion
rate (cf.~Fig.~\ref{cap:star3}). The initial radius of the bubble is assumed
to be smaller than the background dS horizon. \label{cap:star4DS}}
\end{figure}

If the nucleated bubble is smaller than the background horizon, the wall will
implode and form a black hole quite similarly to the situation in the Minkowski
background. The corresponding diagram, Fig.~\ref{cap:star4DS}, is different
from Fig.~\ref{cap:star3} only in the parts relevant to the exterior universe
which is now de Sitter rather than Minkowski (cf.~Fig.~\ref{cap:dsBH} for
the relevant part of the Schwarzschild-de Sitter diagram). 

If the nucleated bubble has super-horizon size, the wall will not collapse but
will asymptotically approach the horizon size, moving inwards. The diagram for
this case is that shown in Fig.~\ref{cap:dS6}.

\begin{figure}
\begin{center}\psfrag{A}{$A$}\psfrag{B}{$B$}\psfrag{C}{$C$}\psfrag{D}{$D$}

\psfrag{E}{$E$}\psfrag{F}{$F$}

\psfrag{E1}{$E'$}\psfrag{F1}{$F'$}\psfrag{F2}{$F''$}

\psfrag{r=0}{$r=0$}\includegraphics[%
  width=2in]{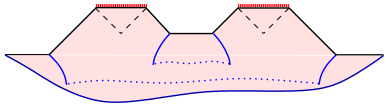}\end{center}

\caption{A future part of the dS spacetime with a nucleated dust-dominated bubble and
a nested dS bubble (cf.~Fig.~\ref{cap:star4DS}). The black hole horizons,
bubble walls, and nucleation surfaces are shown as in Figs.~\ref{cap:dS5}
and \ref{cap:star3}. \label{cap:bu5}}
\end{figure}

Finally, a dust-dominated bubble in a de Sitter background can further nucleate
a nested de Sitter bubble. A diagram for such a spacetime is Fig.~\ref{cap:bu5},
where we have shown both sides of the bubbles rather than only one side as in
Figs.~\ref{cap:star4DS} and \ref{cap:star3}.

\begin{figure}
\begin{center}\includegraphics[%
  width=3in]{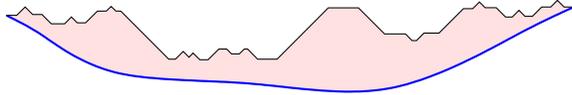}\end{center}

\caption{A sketch of a diagram for the future part of a recycling dS spacetime, featuring
a random arrangement of nested bubbles of all types. Schwarzschild singularities
are not shown. \label{cap:recy}}
\end{figure}

Using all possible nested nucleations as building blocks, we can now sketch
a diagram for a {}``recycling landscape'' featuring a stochastic arrangement
of dS, AdS, and dust-dominated bubbles nucleated randomly within each other
(Fig.~\ref{cap:recy}). The infinite future boundary of the conformal diagram
is a fractal line whose fragments are taken from the diagrams in Figs.~\ref{cap:dsBH},
\ref{cap:dS5}, \ref{cap:dS6}, \ref{cap:dS8}, \ref{cap:star3}, \ref{cap:star4DS},
and \ref{cap:bu5}. Of course, a drawing printed on paper cannot represent the
entire structure of a fractal line. In reality, there are (almost surely) infinitely
many nucleated bubbles between any two such bubbles. 

This diagram concludes the present investigation of the causal structure of
eternally inflating spacetimes.

\section*{Acknowledgments}

The author is grateful to Slava Mukhanov, Matthew Parry, Sergei Solodukhin,
and Alex Vilenkin for stimulating discussions, and to Jaume Garriga and Alex
Vikman for useful comments on the manuscript.

\end{document}